\documentclass[11pt,a4paper]{article}
\usepackage[a4paper, top=3cm, bottom=3cm, left=2cm, right=2cm]{geometry}

\usepackage{helvet}  

\usepackage[pdftex]{graphicx}
\DeclareGraphicsExtensions{.pdf,.jpeg,.png}
\usepackage[dvips]{thumbpdf}
\usepackage{color}

\usepackage{setspace}
\title{Implications of Burn-In Stress  on NBTI Degradation}

\author{Mohd Azman~Abdul Latif,
        Noohul Basheer~Zain Ali, \\
Fawnizu Azmadi~Hussin, and~Mark~Zwolinski}

\date{}

\begin{document}
%

\maketitle

\begin{abstract}
Burn-in is accepted as a way to evaluate ageing effects in an accelerated manner. It has been suggested that burn-in stress may have a significant effect on the Negative Bias Temperature Instability (NBTI) of subthreshold CMOS circuits.  This paper analyses the effect of burn-in on NBTI in the context of a Digital to Analogue Converter (DAC) circuit. Analogue circuits require matched device pairs; NBTI  may cause mismatches and hence circuit failure.   The NBTI degradation observed in the simulation analysis indicates that under severe stress conditions, a significant voltage threshold mismatch in the DAC beyond the design specification of 2 mV limit can result. 
Experimental results confirm the sensitivity of the DAC circuit design to NBTI resulting from burn-in. 
\end{abstract}

\section{Introduction}

In the competitive environment of semiconductor manufacturing, accurate 
reliability prediction results in significant time-to-market and profitability improvements. Prediction quality depends on the manufacturer's ability to characterize process-related instabilities and defects in a given design. 
Burn-in stresses are commonly performed on products to accelerate the fabrication process failure mechanism and to screen out design flaws. 

At sub-micron process technology nodes, it has been suggested that burn-in stress is likely to affect the Negative Bias Temperature Instability (NBTI) \cite{Schroder}, which in turn will affect the operational performance of circuits.

The objective of the work described in this paper is to evaluate the effect of burn-in stress on NBTI, with reference to the performance effect on analogue circuits. A  Digital-to-Analogue Converter (DAC) module was selected as a case study. 
With device reliability models and circuit simulation, this paper analyses the effect of burn-in stress on the shift of key DAC parameters such as the  Integral Non-Linearity (INL), Differential Non-Linearity (DNL) and gain error. 


 

\section{Background}
\subsection{NBTI}
Since the advent of 90nm CMOS technology, NBTI has become  one of the top circuit reliability issues for both PMOS and NMOS devices, because it can severely impact product performance over time. Compared with previous process generations, NMOS hot electron degradation is no longer of such concern. At 45nm, Positive Bias Temperature Instability (PBTI) has an effect on NMOS devices that is about half of that of NBTI on PMOS devices \cite{khan}. 

Several studies have reported on the impact of NBTI on the performance of analogue and digital components.
 It was shown by Kang \textit{et al} \cite{Roy} that the degradation in maximum circuit delay closely follows the trend of threshold-voltage ($V_t$) degradation  in a single PMOS transistor. Their finding was based on a detailed analysis of circuit performance with respect to NBTI degradation, particularly focusing on the maximum delay degradation of random-logic circuits. Kumar \textit{et al} \cite{Sapatnekar} confirmed the effect of NBTI degradation under AC conditions. In addition, Kufluoglu \textit{et al} \cite{Kufluoglu} addressed both PMOS-level measurement delay effects and real-time degradation and recovery by simulation. 
A study performed by Bhardwaj \textit{et al} \cite{Cao} revealed that circuit-level NBTI models can be further improved by considering various process technology-dependent parameters which lead to process variation effects.

 Ball \textit{et al} \cite{Krishnan} have explored the burn-in implications for SRAM circuits. Their approach has demonstrated that the minimum operating voltage, $Vcc_{min}$,  increases during burn-in as a result of NBTI and is of the order of the NBTI-induced $V_t$  shift.

\subsection{Simulation Model}
Schroder and Babcock \cite{Schroder} have thoroughly studied the Time To Failure ($TTF$) relationship to voltage and temperature effects. From their analysis, $TTF$ is affected as follows:

\begin{itemize}\itemsep0pt
\item when the burn-in stress voltage, $V_{stress}$ increases, $TTF$ decreases;

\item when the difference between the nominal voltage, $Vcc$, and $V_t$ increases, $TTF$ decreases; and

\item $TTF$ is inversely proportional to Temperature.
\end{itemize}

The worst case situation is when the system is operated at a high voltage most of the time. However, NBTI degradation can also affect the minimum operating voltage, $Vcc_{min}$, as noted above.

NBTI degradation is less sensitive to $Vcc$ than is NMOS Hot Carrier (NHC) degradation. However, it is more sensitive to temperature \cite{Kimizuka} and occurs even when the transistor is not switching, as long as it is in inversion. 

The following equation shows how the threshold  voltage shift of a PMOS transistor, as a function of the applied voltage and temperature, affects the $TTF$ \cite{Jha}.

\begin{equation} 
TTF(s)=MTTF*f(V_{tp},L)*A
\label{eq1}	
\end{equation}
\begin{equation} 
A= exp(-\gamma*E)*B
\label{eq2}	
\end{equation}
\begin{equation} 
B=exp \left ( \frac {E_{a}}{k}*\left [\frac {1}{T_{j}+273}-\frac{1}{398}\right ]*C \right )
\label{eq3}	
\end{equation}
\begin{equation} 
C=\left [\frac{\Delta{V}_{tp}}{FC}\right ]^\beta
\label{eq4}	
\end{equation}
where
\begin{itemize}
\item$TTF(s)$ is the scaled time to failure in seconds due to voltage and temperature scaling dependencies;
\item $MTTF$ is the mean time to failure at the selected fail criterion (FC);
\item $f(V_{tp}, L)$ is the geometry scaling function for ageing;
\item $\gamma$ is the electric field acceleration factor;
\item $E$ is the electric field  ($V/T_{ox}$) across the gate oxide, of thickness $T_{ox}$;
\item $T_ j$ is the junction temperature;
\item $E_a$ is the thermal activation energy;
\item $FC$ is the $V_t$ shift, defined as the failure criterion for modelling;
\item $\beta$ is a process-dependent variable.
\end{itemize}

This effect can be simulated by applying a signal to the circuit of interest and 
summing the degradation from each time step. 
%
In this case, the effect of the shift in $V_t$, the threshold voltage, for a time varying waveform, can be calculated by using the quasi-static time integral with time in equation  (\ref{eq5}).

\begin{equation} 
TTF=\frac{1}{\frac{1}{t^\prime}*\int{\frac{dt}{TTF(t)}}} = \frac{1}{Avg(1/TTF(t))}
\label{eq5}	
\end{equation}

Lee \textit{et al}  demonstrated the NBTI effect on product reliability degradation \cite{Lee}. 
In addition, their simulator  includes  other reliability mechanisms such as hot carrier injection (HCI) and time-domain-dielectric-breakdown (TDDB) \cite{Lee}. The simulation demonstrates the validity of using a TDDB degradation model to predict the failure rate of a complicated microprocessor. The model is derived using large discrete capacitor/device TDDB data with various temperature, voltage and geometry considerations.

It is noted that even though NBTI degradation occurs under elevated voltage and temperature, the NBTI phenomena show some relaxation. This occurs due to passivation of NBTI-induced silicon dangling bonds by the hydrogen which has diffused from the gate oxide to the interface \cite{Kufluoglu}. There are two types of relaxation that need to be seriously considered for circuit reliability modelling.

\begin{figure}[t]
\centering
\includegraphics[width=\columnwidth]{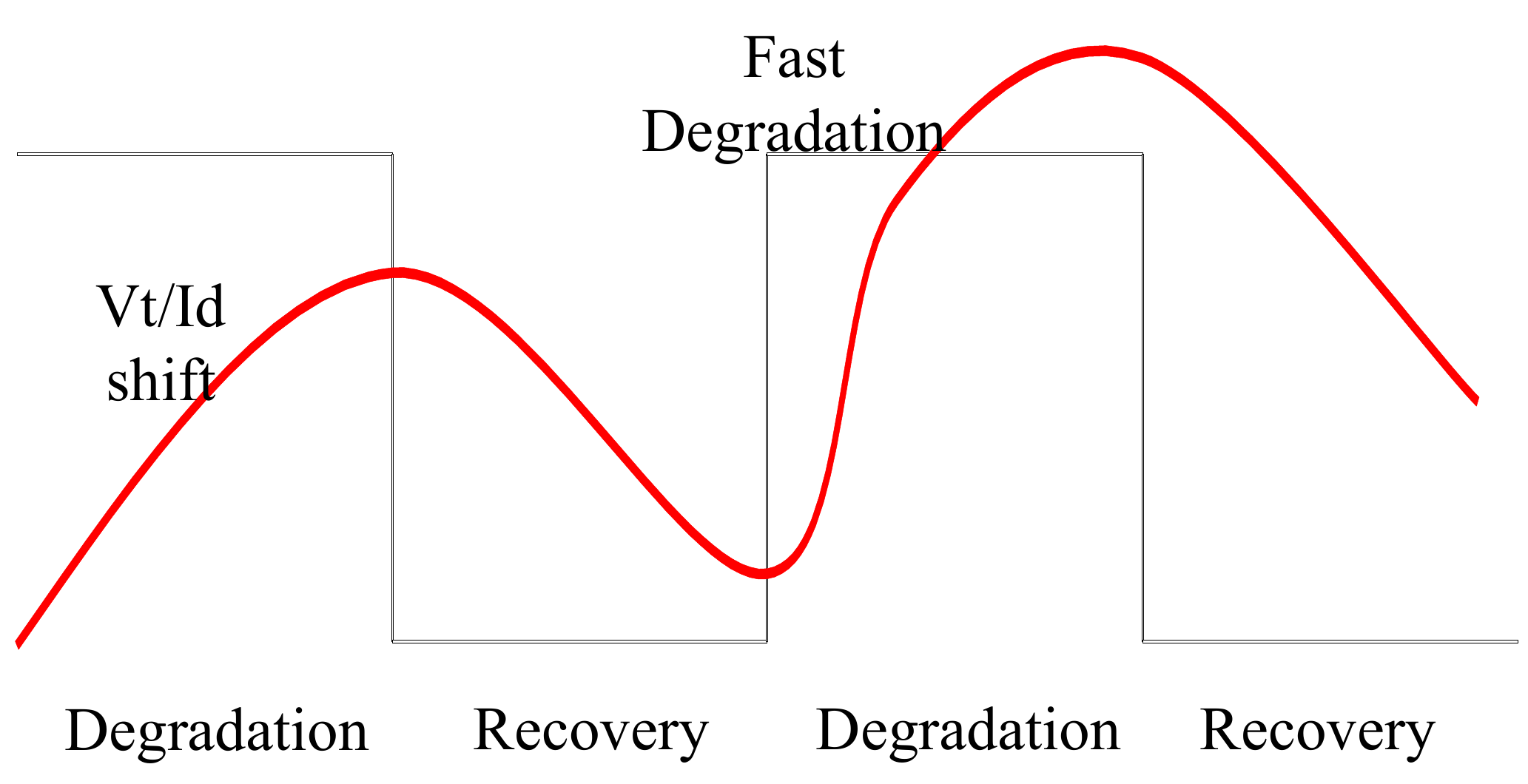}
\caption{NBTI periodic stress and relaxation}
\label{figure 12}
\end{figure}

\begin{enumerate}
\item Fast relaxation: This relaxation occurs as soon as the stress is removed. It is responsible for reduced AC degradation even after accounting for the transistor `ON' time. However, this relaxation mode is not covered in our reliability simulations.
\item Extended relaxation: This relaxation occurs as the device is kept unbiased. Our reliability analysis accounts  for this relaxation mode.

\end{enumerate}
Figure \ref{figure 12} illustrates the two relaxation modes.
Because relaxation lessens the effects of NBTI, a device under continuous usage may suffer a higher degradation than  the reliability simulation predicts. 

\subsection{Assumptions of the reliability simulation model}
We  approximate a complex integrated circuit with a  series failure model. 
We also assume each failure mechanism
has an exponential lifetime distribution. In this way, the failure rate of each failure mechanism is treated as a constant. With these two assumptions, the reliability simulation models, which are often used to extrapolate failure rates, can be validated based on available data.

\section{Reliability Simulation}
\begin{figure}[tbp]
\centering
\includegraphics[width=\columnwidth]{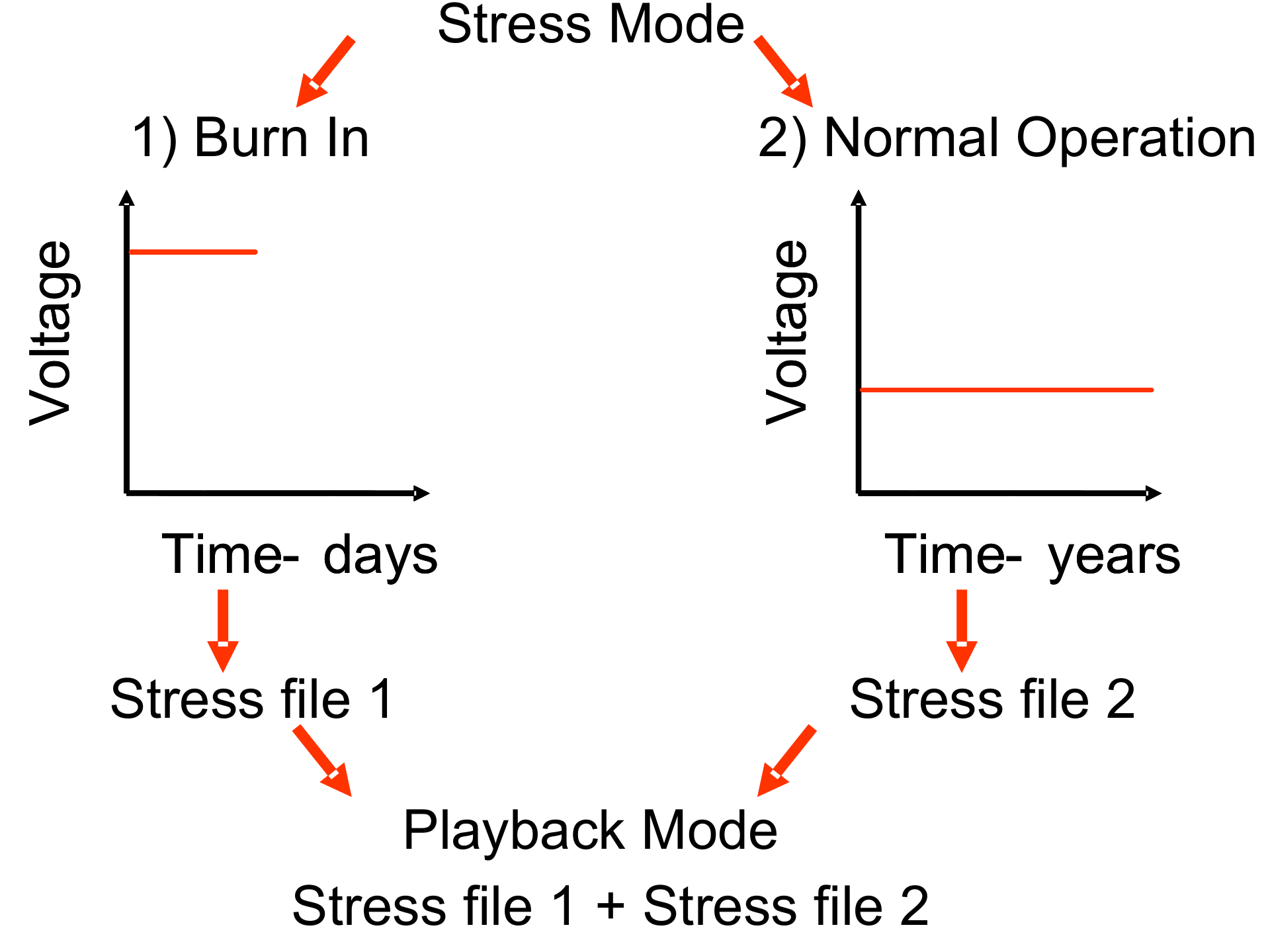}
\caption{Reliability Simulation flow}
\label{fig:aNicePicture}
\end{figure}

For this work,  Intel's internal tool, RELSIM (Reliability Simulation), is used to predict changes in device and circuit performance over a product's lifetime. It further allows simulation of post-degraded circuits to ensure circuit designs meet reliability requirements at end-of-life (EOL) \cite{Latif}. The reliability simulation methodology used in this paper is shown in Figure \ref{fig:aNicePicture}. The simulation has two modes of operation. The first mode, the Stress Mode, calculates the transistor $V_t$ shift. The second mode, the Playback Mode, simulates degraded circuit performance based on Stress Mode results. The simulation is conducted to cover elevated ranges of Process, Voltage and Temperature (PVT). 

The DAC reliability simulation is run in a 3-step process in the design environment.

\begin{enumerate}
\item Simulate the non-degraded behaviour at the typical circuit operating condition ($Vcc$ and temperature).

\item In Stress Mode, calculate the amount of degradation on each transistor.  This is done at a slightly higher voltage and temperature to get a more conservative estimate of the degradation.  

\item In Playback Mode, simulate the degraded circuit, using the degradation calculated in the Stress Mode. 
\end{enumerate}

The Stress Mode is used to report the degradation of a circuit at future times chosen by the user. The user provides the ageing time, the ageing method  (e.g. none, fixed, uniform, bias and temperature user parameters) and a reference degradation value. Also necessary is a degradation parameter file that contains parameters for MOS device stress calculations. During the stress simulation, a stress file is generated at each specified future time. The stress file contains the stressed (degraded) values of each MOS device in the circuit. The degraded circuit values from the initial stress mode can be subsequently used in playback mode. The playback mode produces output signal waveforms for an aged circuit. The information from a stress file is read and a perturbation function is applied to the MOS depending on the degradation model chosen in the stress mode.

The reliability simulator has been used for transistor ageing modelling across major process technologies from 250nm down to 14nm  \cite{Arend}.  The models have been extensively calibrated against  actual silicon test chip data to ensure accuracy \cite{Arend}. The simulator 
can be used to model the minimum $Vcc$ ($Vcc_{min}$) degradation effects. It is able to find the worst case corners,
and takes voltages on all nodes into account. The AC NBTI modelling capability provides more accurate reliability performance predictions than static DC worst-case models. Furthermore, it can be calibrated with the AC circuits to include NBTI recovery, similar to that in \cite{Kufluoglu}.

Another key advantage of this reliability simulator is that the PMOS degradation is modelled with threshold voltage shifts based on  non-uniform I-V degradation. The simulator  models the effect on the MOS transistors I-V characteristics  and the effect on the device parameters and applied voltages. Under the PMOS degradation model, it is suitable for both  digital and analogue simulations.

\section{Case Study : Digital Analogue Converter}

For this case study a video DAC has been used. The performance of the DAC is critical for achieving excellent video quality. The required accuracy of the DAC is based on the differential gain and phase distortion specifications for TV \cite{Wu}. The DAC is designed as a current steering architecture to achieve high accuracy and low distortion of the analogue video signal. The signal range is between zero volts to the maximum nominal analogue video signal swing of 1.3V. The digital input to each DAC is latched on the rising edge of each clock and is converted to an analogue current. For a given digital input, the current source outputs are summed together and directed to the output pin by the differential current switches. An analogue video voltage is created from the DAC output current flowing into the termination resistors. To determine the required output current of the DAC circuit, the video level specifications for the various video formats along with the effective load termination are measured. The LSB output voltage, which ranges between 684 $\mu$V and 1.27 mV, is a function of the supported video format.
Given the circuit mismatch sensitivity of this circuit, paired devices are designed accordingly; typically with greater lengths. 

\begin{figure}[t]
\centering
\includegraphics[width=\columnwidth]{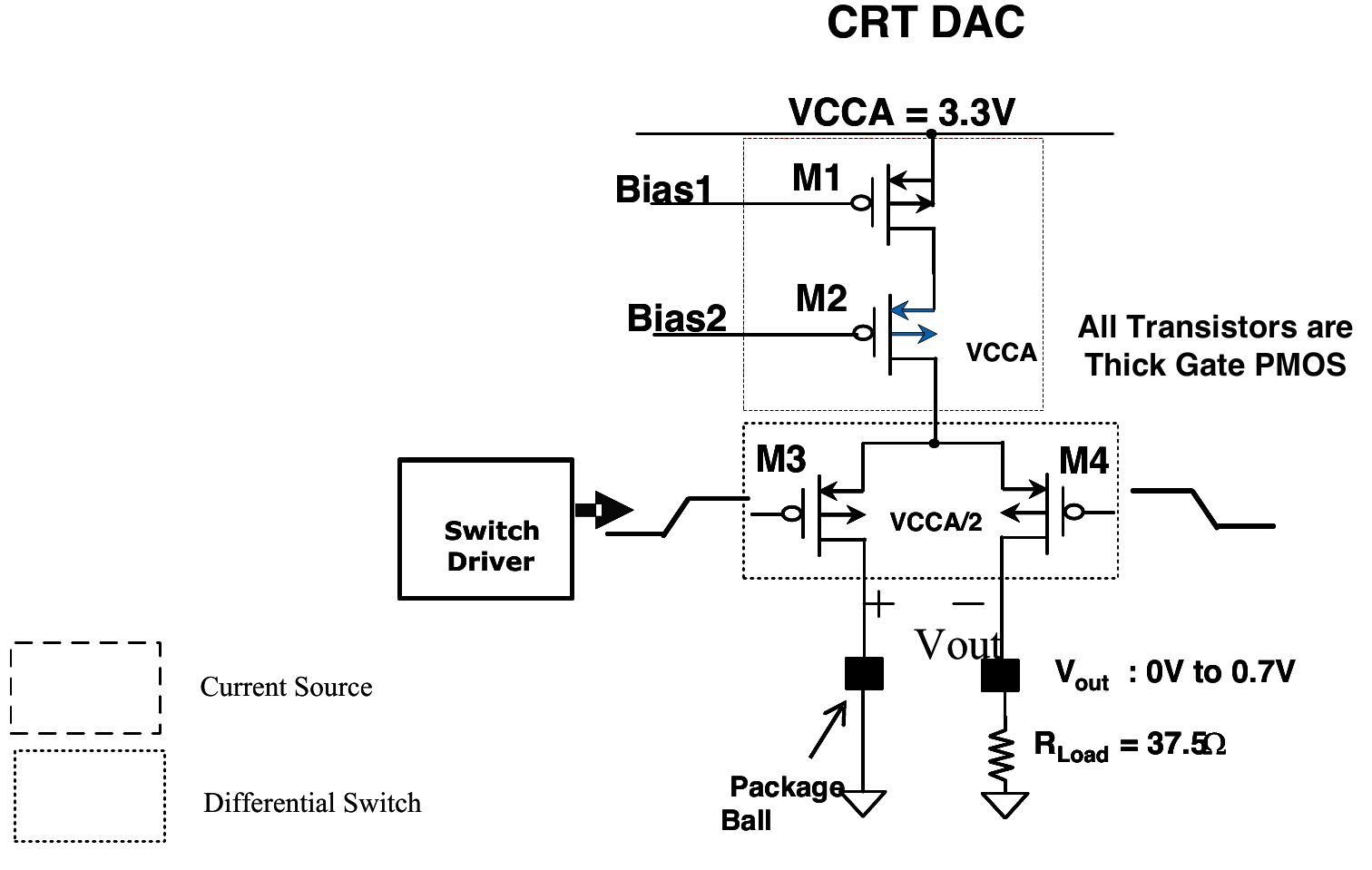}
\caption{Circuit diagram of the current source/differential switch for the CRT DAC}
\label{figure8}
\end{figure}

The DAC  is composed of parallel current switches. This so-called CRT DAC is widely used in  high speed applications specifically for its speed and linearity. The circuit is referenced to an analogue power supply which consists of an array of PMOS current sources and differential current switches (Figure \ref{figure8}).

This DAC operates at 3.3V nominal voltage and implemented in 90nm process technology. It has been shown that the 90nm CRT DAC has sufficient headroom in terms of the circuit performance degradation throughout a 7-year lifetime.

The $V_{t}$ degradation 
is calculated by scaling the gate voltages to the typical analogue operating voltages. The extrapolation is given by equation (\ref{eq8}) \cite{Jha}.
\begin{equation} 
\Delta{V_t} = A*exp(\beta*V_{gs})*exp\left (\frac{E_a}{KT}\right )*t^n
\label{eq8}	
\end{equation}
where:
\begin{itemize}
\item $A$ is the process related pre-factor;
\item $V_t$ is the threshold voltage;
\item $E_a$ is the activation energy in eV ($E_a$ = 0.145 eV was chosen by experiment)
\item $\beta$ is the transconductance parameter ($\beta$ = 0.75 was chosen by experiment);
\item $V_{gs}$ is the gate to source voltage;
\item $K$ is the Boltzmann Constant;
\item $T$ is the temperature in Kelvin;
\item $t$ is the time in years; and
\item $n$ is the voltage acceleration and exponent factor ($n$ = 0.181 was chosen by experiment).
\end{itemize}

\section{DAC ageing simulation and results}

\begin{figure}[t]
\centering
\includegraphics[width=\columnwidth]{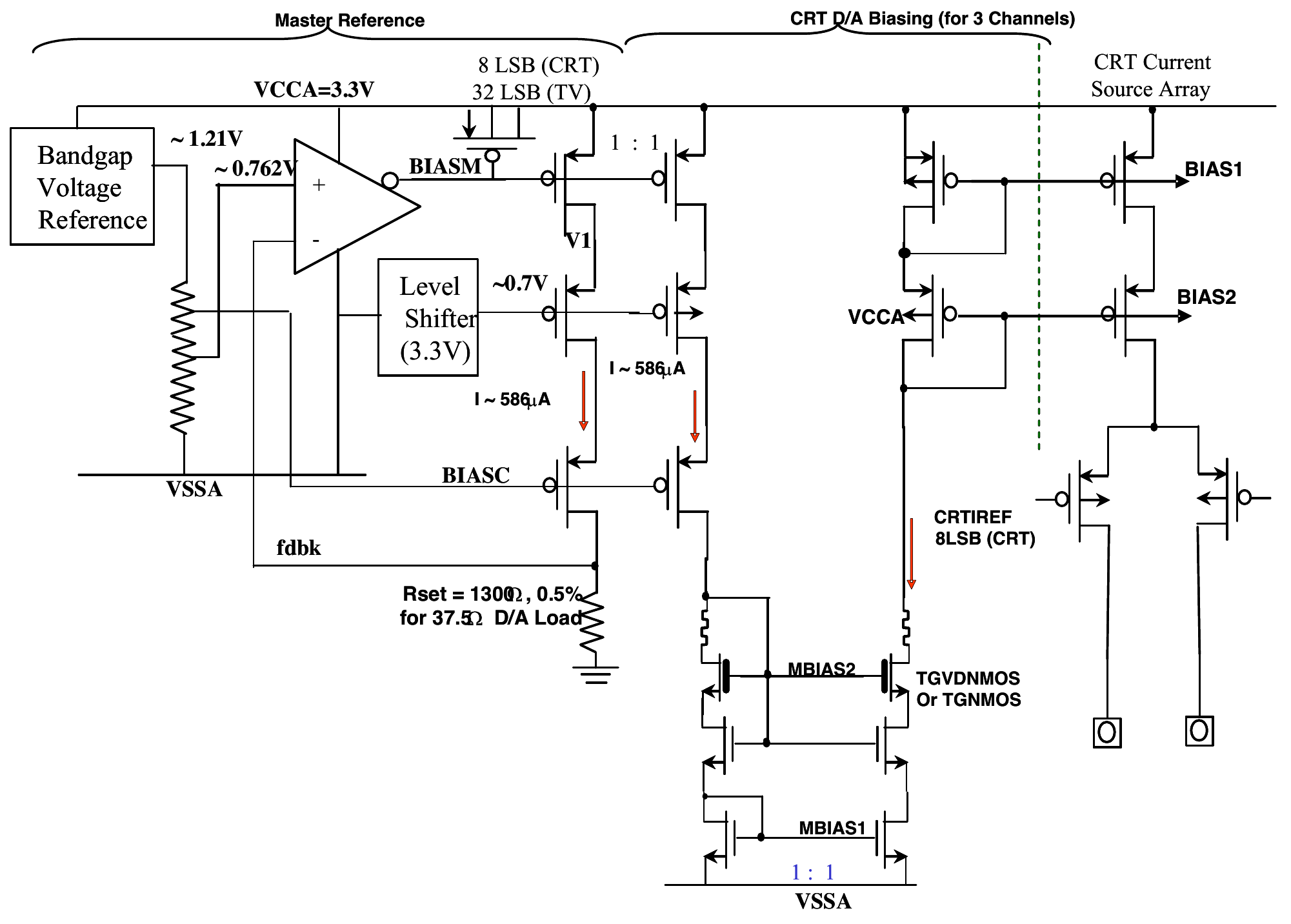}
\caption{Simplified circuit diagram of 8-bit CRT DAC}
\label{figure6}
\end{figure}

For our case study, NBTI analysis was performed on the DAC circuit shown in Figure \ref{figure6}.  We simulated the NBTI behaviour of the DAC  under normal and extreme conditions.

The reliability simulation playback mode analysis was done under the typical corners, for pre-layout schematics with proper loading. For this analysis, the circuit was aged for a 7-year lifetime to check the DAC circuit functionality and the effect of NBTI degradation under burn-in conditions. Table \ref{table1} shows a comparison between three different conditions.

\begin{table}[t]
\centering
\caption{Reliability simulation parameters across three different conditions}
\label{table1}
\begin{tabular}{|c|c|c|c|} \hline
Parameters & Fresh & Burn In (Stressing Mode) & Age (Playback Mode)\\
\hline
Stress Skew & Typical Corner & Typical Corner & Typical Corner \\
Stress Voltage & 3.3V & 4.6V & 3.3V \\
Temperature & 100$^\circ$C & 110$^\circ$C & 110$^\circ$C\\
Use time & Time 0 & 168 hrs & 7 years \\
\hline
\end{tabular}

\end{table}

We analysed the matched devices in Figure \ref{figure6} and observed slightly different degradation behaviours. There are two key device parameters that are critical to degradation behaviours.

\begin{enumerate}
\item Drain current, $I_{d}$: The current reaches its maximum value and maintains that value for higher drain-to-source voltages. A depletion layer located at the drain end of the gate accommodates the additional drain-to-source voltage. This behaviour is referred to as drain current saturation, $I_{dsat}$. Drain current saturation therefore occurs when the drain-to-source voltage equals the gate-to-source voltage minus the threshold voltage. 
\item Threshold voltage, $V_t$: A group of transistors has a Gaussian profile about a mean. Experimentally, it has been shown that the difference in threshold voltages between 2 identically sized transistors behaves as described in equation \ref{eq7} \cite{Schroder}.

\begin{equation} 
\sigma_{\Delta} = \frac{A_{V_t}}{\sqrt{WL}}
\label{eq7}	
\end{equation}

where $A_{V_{t}}$ is a technology conversion constant (in mV$\mu$m), and WL denotes the product of the transistor's active area.

\end{enumerate}

From the simulation results, it is observed that the $I_{dsat}$ and the $V_t$ degradations of both the matched devices  M1 and M2 pairs and the differential switch M3 and M4 pairs at 3.3V nominal condition of VCCA were comparable. The $I_{dsat}$ degradation at 3.3V for all transistors seems greater than that of the 4.6V burn-in. However, the delta of 3.3V and 4.6V is not enough to conclude that degradation happens even at 3.3V nominal. On the other hand, the $V_t$ degradation shows a significant difference at the two voltage readings. At the 4.6V burn-in condition, the degradation of the matched devices  in both M1 and M2 pairs gives a $V_t$ mismatch of 5.2 mV, compared with the DAC specification of 2 mV,  as shown in Table \ref{table2}. These two current source pairs have a higher $V_t$ mismatch compared to the differential pairs. This mismatch may cause the DAC to malfunction. 

\begin{table}[t]
\centering
\caption{Reliability Simulation result comparing current source/differential pairs at 3.3V and 4.6V}
\label{table2}

Reliability Simulation - Burn In mode @ 3.3V (V nominal)
\begin{tabular}{|c|c|c|c|} \hline
Transistor &Pair &idsat (\%) & Vt mismatch (mV)\\
\hline
M1 & Current Source & 2.157 & 0.902\\
M2 & Current Source & 2.157 & 0.902\\
M3 & Differential & 2.047 & 0.731\\
M4 & Differential & 2.047 & 0.731\\
\hline
\end{tabular}

Reliability Simulation - Burn In mode @ 4.6V (1.4X V nominal)
\begin{tabular}{|c|c|c|c|} \hline
Transistor & Pair & idsat (\%) & Vt mismatch (mV)\\
\hline
M1 & Current Source & 1.905 & 5.239\\
M2 & Current Source & 1.905 & 5.239\\
M3 & Differential & 2.039 & 2.059\\
M4 & Differential & 2.039 & 2.059\\
\hline
\end{tabular}

\end{table}


For simulation purposes, the DAC performance data in Table \ref{table 13} was simulated at 3.3V nominal as well as at 4.6V at elevated voltage. The performance data focused on the critical parameters: 
\begin{enumerate}
\item Differential Non-Linearity (DNL)
\item Integral Non-Linearity (INL)
\item Gain Error
\item Offset Error
\item Output Current
\end{enumerate}
From the simulation results, the DAC key performance parameters were generally within the specification criteria, by percentage change. However, it is noted that the gain error measurement of 2.68V at 4.6V is equivalent to a 7.2\% change, slightly higher than the +/-5\% specification limit. 

\begin{table}[t]
\centering
\caption{Simulation data of DAC parameters at 3.3V and 4.6V}
\label{table 13}
\begin{tabular}{|c|c|c|c|c|} \hline
DAC Parameters & Spec & Sim\#1 @ 3.3V & Sim\#2 @ 4.6V & Percent Change\\
\hline
DNL (20MHz)& +/-1 LSB& 0.1734 LSB& 0.1803 LSB& 3.98\%\\
INL (20MHz)& +/-1 LSB&0.0387 LSB&0.0417 LSB&7.75\%\\
Gain Error&+/-5\%&2.500 V&2.68 V&7.20\%\\
Offset Error&+/-5\%&0.000387 V&0.000403 V&4.13\%\\
Output Current&0-21 mA &4.58 mA&4.93 mA&7.64\%\\
\hline
\end{tabular}
\end{table}

%

\section{DAC burn-in experiment}

For the burn-in experiment, the voltage supply of interest, the analogue CRT DAC power supply, VCCA, was elevated to a burn-in voltage of 4.6V. The nominal voltage for this power supply is 3.3V.  Three hundred units from three fabrication lots were analysed.

These units were subjected to burn-in stress for 30 minutes followed by post burn-in checkpoint (PBIC). All units completed a cumulative 168-hour burn-in. Of 300 experimental units, 30 good units were sampled for specific DAC burn-in characterization. From these 30 samples, 1 unit from each of the three fabrication lots was marked as the control unit. These control units were tested first and used for reference, while the rest of the units were tested after each of the burn-in phases at a temperature of 115$^\circ$C.

ATE testers were used to take a series of electrical measurements (timing and parametric shift)   at different phases: pre-burn-in, time zero, and follow-on (0.5 hours, 12 hours, and 168 hours burn-in). The same five critical performance parameters as in the simulation were measured. Two sets of data were collected: at the nominal voltage of 3.3V; and at the elevated burn-in voltage of 4.6V.  

\section{DAC burn-in results and discussion}

\begin{figure}[t]
\centering
\includegraphics[width=\columnwidth]{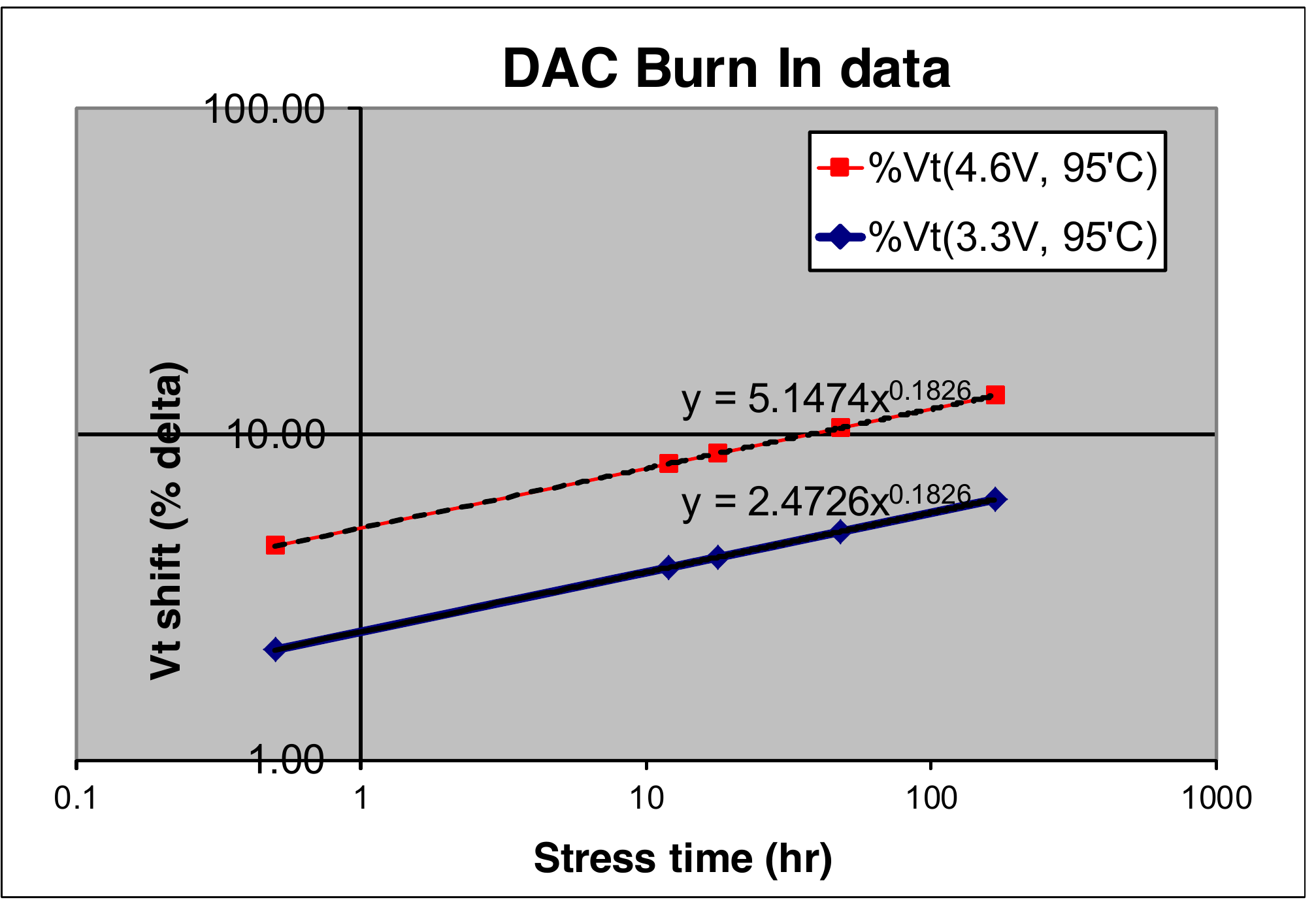}

\caption{DAC Burn In comparison: 3.3V (nominal) vs 4.6V (stress) supply voltage with respect to stress time}
\label{figure 10}
\end{figure}

We present in Figure \ref{figure 10} experimental data from stressing discrete transistors that illustrates the increase in $V_t$ with respect to the NBTI stress time. The graph was plotted by applying the power law in equation (\ref{eq9}).

\begin{equation} 
\label{eq9}
\Delta V_{t} = A*t^n
\label{eq77}	
\end{equation}

Figure \ref{figure 10} compares the results from time zero at 3.3V with those at 4.6V elevated voltage with respect to the stress time.
It is apparent that NBTI has a power-law time dependence. When plotted on a log-log scale, we see that a higher voltage difference between gate and source will result in a higher degradation. This shows that $\Delta V_t$ has a power-law dependence with respect to time. As a result, the curve for the 4.6V stress (black) after burn-in is shifted upwards compared to the curve in blue at 3.3V,  which represents the time zero stress. 

The slope of this line is called the `power-law slope', $n$. It is technology dependent and typically ranges in value from 0.15 to 0.30 \cite{Schroder}. The low value of n (n $<$ 1) gives rise to the `quasi-saturating' behaviour. It is important to note that NBTI degradation has been shown to follow a power law time dependence due to the physics of the electrochemical reaction/diffusion reaction underlying NBTI degradation \cite{Schroder}. Therefore, the functional form is not merely a curve-fitting exercise but rather a necessary consequence of the degradation physics.

We measured the same five key DAC parameters as in the simulation. These measurements were taken after the last burn-in readout of 168-hours under a severe stress condition at 4.6V. A 168 hour burn-in at 95\% is equivalent to 7 years of normal operating condition, as in the reliability simulation. The key test results measured before and after burn-In stress are summarized in Table \ref{table3}. The changes in the parameters after the burn-in are comparable to those resulting from the reliability simulation, Table \ref{table 13}. We can therefore be reasonably confident that the ageing simulation and the emulation of ageing through burn-in are consistent. In other words, we have a strong correlation, but we cannot prove causality.

Based on the test results measured, the gain error measurement after the post burn-in stress was the only parameter that has a significant spike as compared to the rest of the key parameters. A significant increase in the gain error of 43.5$\%$ was observed. The ideal situation is that the DAC's gain error has to be zero. The gain measurement was collected on the ATE tester. The pre burn-in is assumed to be an ideal case with zero gain error. Table \ref{table4} shows that as the DAC input code increases, the output voltage increases accordingly to 2.5V (V$_{\rm ref}$). 

\begin{table}[t]
\centering
\caption{The Vout data taken before and after Burn-In stress}
\label{table4}
\begin{tabular}{|c|c|c|}\hline
Input code&Vout @ pre Burn-In&Vout @ post Burn-In\\
\hline
0&0.000&0.000\\
1&0.167&0.239\\
2&0.333&0.478\\
3&0.500&0.717\\
4&0.667&0.956\\
5&0.833&1.195\\
6&1.000&1.434\\
7&1.167&1.673\\
8&1.333&1.912\\
9&1.500&2.151\\
10&1.667&2.390\\
11&1.833&2.629\\
12&2.000&2.868\\
13&2.167&3.107\\
14&2.333&3.346\\
15&2.500&3.585\\
\hline
\end{tabular}

\end{table}

\begin{figure}[t]
\centering
\includegraphics[width=\columnwidth]{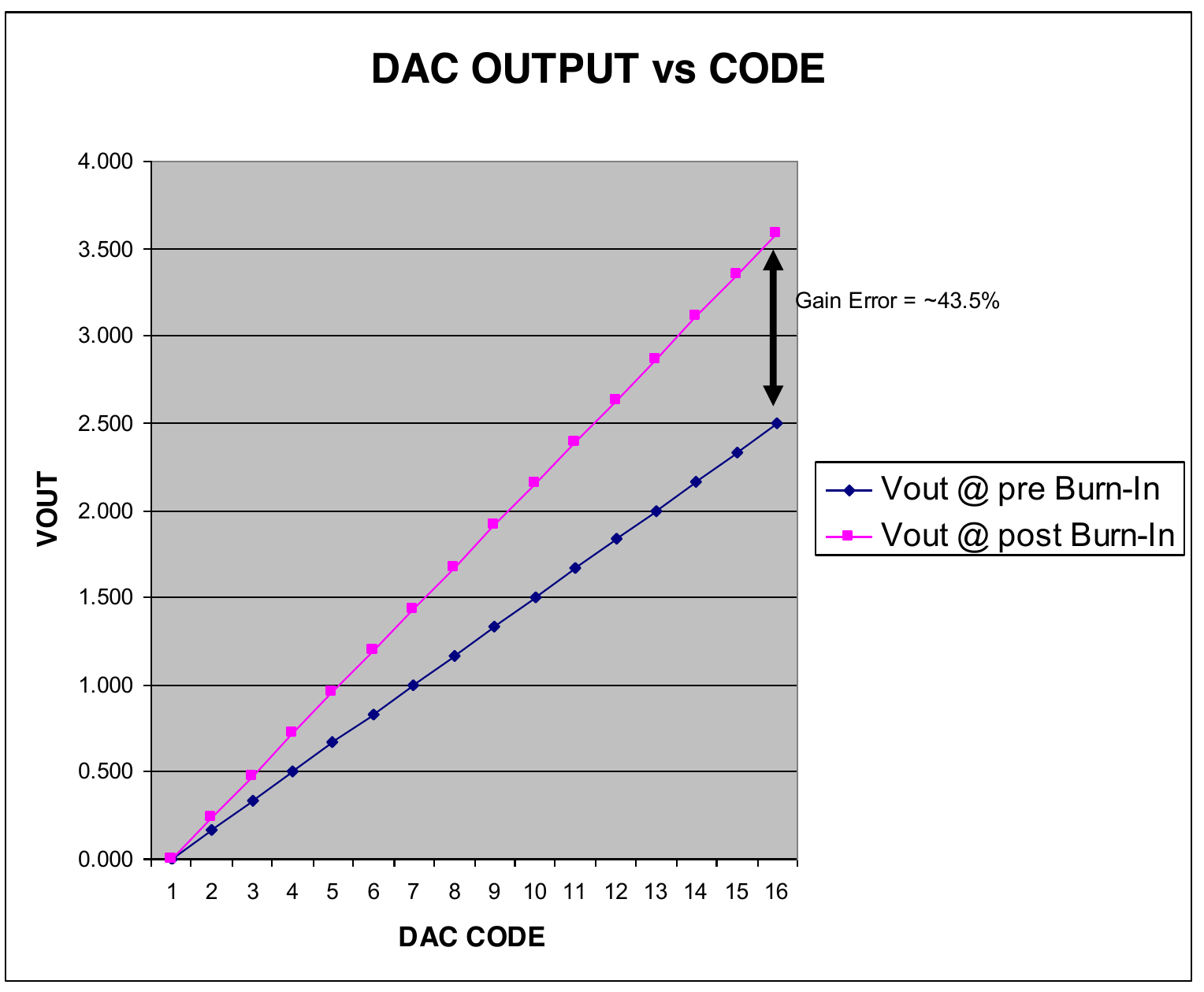}
\caption{DAC characteristic showing an excessive gain error of 43.5$\%$}
\label{figure66}
\end{figure}

The post burn-in Vout was measured at the ATE tester and the Vout values increased by $\sim$43.5$\%$ compared to the typical gain error percentage of less than $\sim$20$\%$. The gain was calculated as A$_{\rm v}$= 3.585/2.5 = $\approx$1.43406667. Hence, the gain error percentage is $\approx$43.5$\%$.
Figure \ref{figure66} shows the DAC transfer functions of ideal (pre burn-in) vs actual DAC (post burn-in). 

In this specific case, the gain error has created a span greater than the desired ideal case. The transfer function is modelled as a typical straight line as commonly described by \textit{y = mx + c} equation, where:

\begin{itemize}
\item \textit{y} is the output of the DAC

\item  \textit{m} is the slope of the transfer function

\item  \textit{x} is the input of the DAC

\item  \textit{c} is the offset voltage
\end{itemize}

Typically, an ideal DAC has a gain, \textit{m}, of 1 and an offset, \textit{c}, of 0 and hence the output tracks the input in a precise linear manner. However, for the real DAC, it has non-ideal gain and offset values which normally can be compensated once the values are determined.

For the data taken in this burn-in experiment, the design is based on an 8-bit DAC with a 0V to 2.5V nominal output span. When the digital input is set to a full scale, a 3.585V output is measured.

From this data, the actual gain error, measured in percentage, can be determined by multiplying the output voltage at post burn-in by the output voltage at pre burn-in, 3.585V/2.5V = $\approx$1.43406667. The gain error is calculated with the assumption that the offset error is zero while the span error is measured at 850 mV, giving an actual span of 3.585V.

Notice, however, the significant increase in the gain error of 43.5$\%$. In other words, the standard reliability simulation tool correctly predicts changes in key DAC parameters, including the effect of NBTI, but appears to underestimate the change in the gain error. It is well-known that mismatch between current mirror paired devices will cause such gain errors \cite{Agostinelli}. Therefore, it is reasonable to conclude that the increase in gain error after burn-in has resulted from transistor pairs becoming mismatched, and such a mismatch is likely to be due to changes in the threshold voltages. As has been shown, above, NBTI causes significant changes in $V_t$. Therefore it is again reasonable to conclude that the change in gain error, as a result of burn-in, has been caused by NBTI.

\begin{table}[t]
\centering
\caption{Summary of 168 hours Burn-In experiment of DAC}
\label{table3}
\begin{tabular}{|l|c|c|c|}\hline
Items&Pre Burn-In&Post Burn-In&Percent Change\\
\hline
DNL mean&0.185 LSB&0.197 LSB&6.48\%\\
INL mean&0.235 LSB&0.246 LSB&4.68\%\\
Gain Error&2.500V&3.585V&43.50\%\\
Offset Error&0.000344V&0.000361V&4.90\%\\
Output Current&5.9 mA&6.02 mA&2\%\\
\hline
\end{tabular}

\end{table}

\section{Conclusion}
The NBTI degradation observed in the reliability simulation of a DAC circuit revealed that under a severe stress condition such as a 40\% increase in the nominal voltage supply, a significant voltage threshold mismatch, beyond the 2 mV limit, was recorded. A burn-in experiment on the DAC circuit was performed to verify the simulation. A correlation between the simulation results and the burn-in behaviour was observed, but the change in the gain error was significantly greater than predicted. 

%
%


\end{document}